\documentclass[aps,prr,amsmath,amssymb,floatfix,twocolumn]{revtex4-1}

\usepackage{multirow}
\usepackage{bbold}
\usepackage{graphicx}
\usepackage{subfigure}
\usepackage{color}
\usepackage{hyperref}
\usepackage[justification=raggedright]{caption}
\usepackage{filecontents}

\begin{document}
\preprint{APS/123-QED}

\title{Simulating alternating bias assisted annealing of amorphous oxide tunnel junctions}
\author{Alexander C. Tyner$^{1,2}$}
\email{alexander.tyner@su.se}
%\author{??}
\author{Alexander V. Balatsky$^{1,2}$}
\email{alexander.balatsky@uconn.edu}

\affiliation{$^1$ Department of Physics, Institute for Materials Science, University of Connecticut, Storrs, Connecticut 06269, USA}

\affiliation{$^{2}$ Nordita, KTH Royal Institute of Technology and Stockholm University 106 91 Stockholm, Sweden}

\date{\today}

\begin{abstract} 
\noindent 
Amorphous oxide tunneling barriers, primarily formed from aluminum, represent one of the most widely adopted platforms for superconducting quantum bits (qubits). To overcome challenges associated with defects and sample variance among the tunneling barriers, the methodology of alternating bias assisted annealing (ABAA) was introduced in Pappas et. al\cite{pappas2024alternating}. The process of applying alternating bias to the barrier and subsequently aging before use was shown to reduce defects in the barrier. Namely, defects that give rise to two-level systems, coupling to the qubit and expediting decoherence. In this work we replicate an expedited ABAA process through a combination of ab-initio molecular dynamics and machine-learned potentials, illuminating how ABAA effects the energy landscape of the barrier. 
\end{abstract}

\maketitle
\par 
\section{Introduction} 
Amorphous oxide tunneling barriers serve as the building blocks for modern superconducting quantum bits, generally referred to as qubits\cite{nakamura1999coherent,devoret1995quantum,kjaergaard2020superconducting,krantz2019quantum,bravyi2022future}. Quantum computers utilizing such qubits have realized immense progress in recent years, growing both in scale and coherence times. Nevertheless, a primary limitation to coherence time in these platforms remains the presence of defects in the amorphous oxides tunneling barrier which give rise to two-level systems\cite{siddiqi2021engineering,burnett2019decoherence,lisenfeld2015observation,lisenfeld2019electric}. In many cases the energetic splitting of the two-level system is commensurate with the frequency of the qubit, allowing the two to couple and expedite decoherence. 
\par
\begin{figure}
    \centering
    \includegraphics[width=8cm]{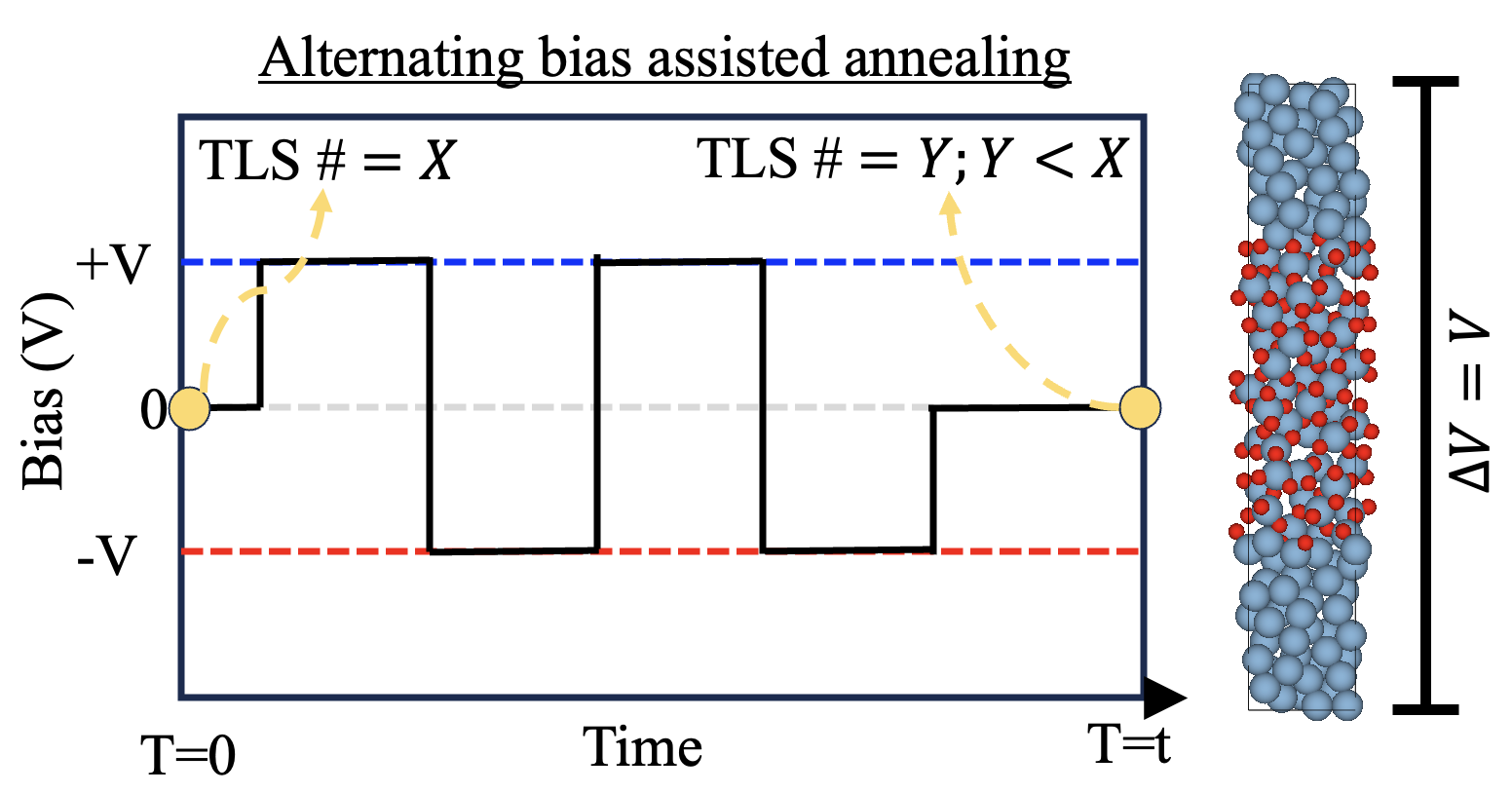}
    \caption{Schematic of alternating bias assisted annealing (ABAA) protocol, wherein an electrical bias of alternating magnitude is applied to the amorphous oxide tunneling barrier, depicted on the right. The number of two-level systems (TLSs) in the qubit frequency range before ABAA, labeled X, has been observed to decrease relative to the TLS number following ABAA, labeled Y in experimental studies\cite{pappas2024alternating}. }
    \label{fig:Fig1}
\end{figure}

 
Recently, Pappas et. al\cite{pappas2024alternating} proposed a protocol to reduce the number of TLSs and increase coherence times referred to as alternating bias assisted annealing (ABAA). In this method a series of alternative electric biases are applied to the amorphous oxide barrier, which are ubiquitously composed of amorphous aluminum oxide (a-AlO$_{x}$). Following application of the alternating bias potential, the tunneling barrier is allowed to age (relax), prior to use in a superconducting circuit. The authors found that this process consistently reduced the number of identifiable TLSs and increased coherence times. A schematic of this process is given in Fig. \eqref{fig:Fig1}.

\par 
A leading hypothesis for the efficacy of this protocol is that application of the alternating bias provides a perturbation to shift the system away from a local energy minima and explore the energy landscape, locating a more stable configuration with fewer defects. Minimization of defects in turn reduces the number of TLSs. While there exist a number of reasonable hypotheses for the microscopic origin of TLSs in an amorphous oxides, one possibility to illuminate the connection between structural defects and TLSs is that dangling bond features in the oxide support an additional rotational degree of freedom (in addition to transverse and longitudinal modes)\cite{tyner2025identification,heath2025many,varley2023dangling,wang2024superconducting,muller2019towards}. This additional degree of freedom gives rise to an anharmonic well\cite{varma2023theory} which can be identified as the TLS, as illustrated in Fig. \eqref{fig:Fig2}.
\par 
Computational expense associated with the tunneling barrier size as well as the time-scale of the ABAA protocol have thus far discouraged computational modeling of this method. In this work we utilize a combination of \emph{ab-initio} computations and machine-learned interatomic potentials to simulate and analyze the ABAA method. Our results shed light on the necessary time-scales for this procedure as well as how this protocol allows for a traversal of the energetic landscape of the tunneling barrier to identify new energetic minima, in agreement with the experimental findings. We furthermore comment on the effects of temperature and field strength, two degrees of freedom which have been explored experimentally in recent works\cite{iaia2025non}. 

\section{Simulating ABAA}
The starting point for computational simulation of ABAA is construction of the tunneling barrier, primarily the amorphous oxide. To computationally generate the amorphous aluminum oxide we follow a melt and anneal protocol first introduced in Ref. \cite{cooper2000density} for silicon oxides and repeated in Ref. \cite{tyner2025identification} for aluminum and tantalum oxides. Full details of this process can be found in Ref. \cite{tyner2025identification}  but is summarized here for clarity. 

\begin{figure}
    \centering
    \includegraphics[width=8cm]{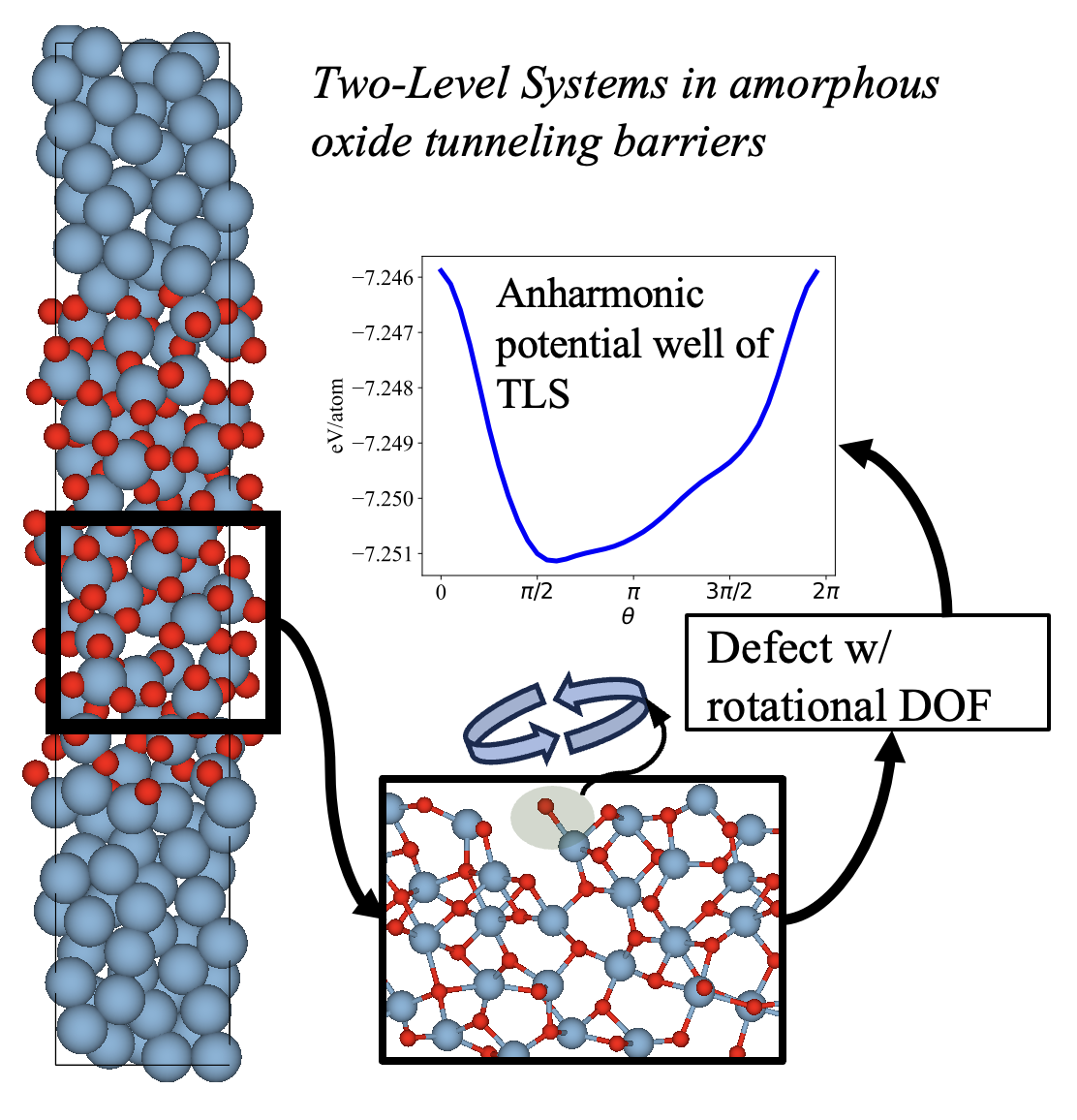}
    \caption{Left: Amorphous aluminum oxide tunneling barrier with Al atoms in blue and O atoms in red. Bottom: Enhanced view of oxide where an oxygen atom can be seen to exhibit the features of a dangling bond, isolated in real-space. The spatial isolation of this atom gives rise to a rotational degree of freedom. Recomputing the energy along a path in this rotational degree of freedom yields the anharmonic potential well shown in the top right. These wells can give rise to two-level systems which couple to the qubit, expediting decoherence.  }
    \label{fig:Fig2}
\end{figure}
\par 
\emph{Generation of the amorphous oxide structure: } We begin with a crystalline, stoichiometric sample of Al$_{2}$O$_{3}$ containing 120 total atoms.  To form the amorphous structures we utilize Car-Parinello molecular dynamics (CPMD)\cite{hutter2012car} as implemented within the Quantum Espresso software package\cite{QE-2020,Perdew1996}. First, the atoms are randomly displaced from their starting positions by a maximum of 2 percent of the lattice vector magnitude in each direction. Throughout the CPMD computations we use a $5 \times 5 \times 5$ real-space mesh for the wavefunction and charge density FFT. A plane-wave cutoff of 60 Ry, a timestep of 5.5 a.u. (1 a.u.$= 2.4189 \times 10^{-17} s$), a fictitious electron mass of 300 a.u. (1 a.u. = $9.10939 \times 10^{-31}$kg)\cite{cooper2000density} and non-relativistic pseudopotentials from the {\color{black}PseudoDojo}~\cite{van2018pseudodojo} data base are used.

\par 
After the atomic positions are randomized the structures are melted by rescaling the velocities to reach a temperature of $4000\pm 400 K$, which is above the melting point of $2345 K$\cite{schneider1967effect}, for 200 time steps. In this period the system becomes liquid-like. We then apply Nose-Hoover thermostats to the ions and electrons. The ion thermostat is set to $2100 K$ and a characteristic frequency of 15 THz, determined by the peak in the crystalline phonon density of states. For the electrons the thermostat was set to a target kinetic energy of 0.0075 a.u. and a characteristic frequency of 1350 THz. The liquid phase is equilibrated for 2800 time steps within the canonical ensemble. We then cool this sample at a rate of $2.4 \times 10^{15} K s^{-1}$. At this rate the sample is quenched from $2100 K$ to $300 K$ in $\sim 0.7 ps$. To ensure stability, the amorphous structure is then put through a secondary annealing cycle in which the temperature is raised to $900 K$ over the course of $0.1 ps$, equilibrated at this temperature for $0.53 ps$ and then cooled to $300K$ over $0.25 ps$. Finally, the system is allowed to equilibrate at $300K$ for 3500 time steps. The final structure is then allowed to evolve within an NVE ensemble and does not deviate significantly in temperature after 3500 time steps indicating stability. 
\par 
\emph{Generation of the tunneling barrier: } Following generation of the amorphous oxide, we create an interface on each end to a 6 atomic layer slab of $(111)$-oriented aluminum. The cell and atomic positions of the full Al-$a$-Al$_{2}$O$_{3}$-Al barrier are then relaxed within the context of density functional theory. The resulting structure is shown on the left in Fig. \eqref{fig:Fig2}. We note that this structure is distinct from those constructed in Refs. \cite{dubois2016constructing,cyster2021simulating,kim2020density} due to the method of construction. Namely, we have chosen to form the oxide prior to creation of an interface, whereas Refs. \cite{dubois2016constructing,cyster2021simulating,kim2020density} pursue the computationally expensive approach of replicating atomic deposition on an aluminum thin film. As such the interfaces for the structure simulated in this work resemble the top interface of the structures in Ref. \cite{cyster2021simulating} in which aluminum is deposited on the amorphous oxide, rather than the bottom interface in which oxygen is adsorbed on the surface of aluminum. 

\begin{figure}
    \centering
    \includegraphics[width=9cm]{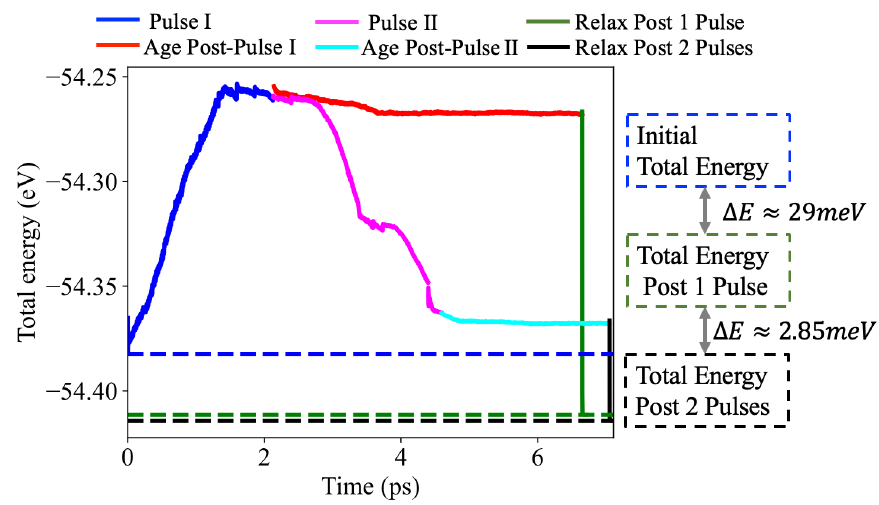}
    \caption{Total energy of the amorphous oxide tunneling barrier as a function of time at 300K. The molecular dynamics simulation begins by applying a positive bias (Pulse I) to the barrier after it has been relaxed within density functional theory. Pulse I is terminated when the energy plateaus at $\sim 2ps$, after which two subsequent simulations are initiated. In the first, the system is allowed to age in the absence of the bias simulating aging, shown in red. Following this process the structure is then relaxed at 0K to compare the final and initial energy, shown in green. In the second, a bias of opposite sign and equal magnitude is applied, shown in magenta. The bias is again terminated when the energy begins to plateau and allowed to age in the absence of an applied field, shown in cyan. Following this process the structure is then relaxed at 0K to compare the final and initial energy, shown in black. Each pulse is found to lower the total energy of the system upon aging and relaxation.}
    \label{fig:Fig3}
\end{figure}
\par 
\emph{Simulation of alternating bias: } In order to simulate the application of an alternating bias on the tunneling barrier we again utilize Car-Parinello molecular dynamics within the Quantum Espresso software package\cite{QE-2020,Perdew1996}. We utilize the same fictitious electronic mass, energy cutoff and time step used in generation of the amorphous oxide. We choose this methodology as it allows for inclusion of an applied electric field and high accuracy without the computational expense of fully self-consistent MD-DFT. Computational efficiency is vital as experimental ABAA protocols involve cycle on the order of many seconds. This time-scale is unattainable within our computational simulations where each time step represents $\sim1.3\times 10^{-4} \;ps$. However, within the simulation it is possible to achieve timescales of 1-10 picosconds. We estimate that this sufficient to gather actionable information given that this time scale is commensurate with the high-end of TLS frequencies, and the qubit frequency should set the maximal necessary simulation timescale.
\par 
Prior to inclusion of the electric field we ensure that the electrons are located on the Born-Oppenheimer surface by relaxing within a Car-Parinello simulation at 0K using damped dynamics to propagate the electronic degrees of freedom. An electric field is then included in the simulation through the modern theory of polarization. The potential difference along the $c$-axis is set to 0.5V. The ionic and electronic degrees of freedom are then allowed to evolve in an NVT ensemble, fixing the temperature to $30$K. Throughout this simulation the total energy of the system is tracked, as shown in Fig. \eqref{fig:Fig3}. In this figure we observe that application of the electric field increases the energy of the system during the first pulse. This is reasonable as the potential bias exerts a force on the atoms shifting them away from a energetic minima reached during the prior DFT relaxation. Following more than 10,000 time-steps, the total energy of the system is seen to plateau. This plateau suggests that the system has reached a new equilibrium state. 
\par 
At this point there are two possibilities for proceeding with the simulation: (A) turn off the electric field and allow the system to continue to evolve (also referred to as age) in this equilibrium state; (B) alternate the bias and restart the simulation. We will first consider option (A), setting the electric field strength to zero and continuing the NVT ensemble evolution. The energy of the system during this process is visible in Fig. \eqref{fig:Fig3}. We note that the total energy remains generally constant before slowly decreasing as the system locates a stable equilibrium in the absence of the applied field. From this equilibrium, to compare the stability of the final state with the initial state we must bring the system back to the Born-Oppenheimer surface at $0$K. To do so we restart the simulation following the aging cycle using damped dynamics to propagate the ionic and electronic degrees of freedom. The resulting energy of the system is marked by the blue dashed line in Fig. \eqref{fig:Fig3} and falls below the initial energy, indicating that the application of a bias followed by aging has allowed the system to realize a new energetic minima with enhanced stability compared to the initial system. 
\par 
Next we consider option (B): alternate the bias and restart the simulation. This amounts simply to restarting the simulation with an electric field of the same magnitude and opposite sign. The system is then allowed to evolve in the NVT ensemble. Tracking the total energy, shown in magenta in Fig. \eqref{fig:Fig3}, we observe the total energy decrease as the system appears to reverse the energetic pathway of the first pulse before reaching a new equilibrium at which point the electric field is turned off and the system is allows to age, shown in cyan in Fig. \eqref{fig:Fig3}. In this aging cycle we note that the energy remains mostly constant, again indicating that the system has reached a stable equilibrium from which a relaxation to return the system to the Born-Oppenheimer surface can be performed and the energy compared with the initial state. This relaxation is shown in black, with the final energy marked by a horizontal black dashed line demonstrating that the final state has reached an energetic minima falling below both the initial state and that reached after a single pulse. 
\par
\begin{figure}
    \centering
    \includegraphics[width=8cm]{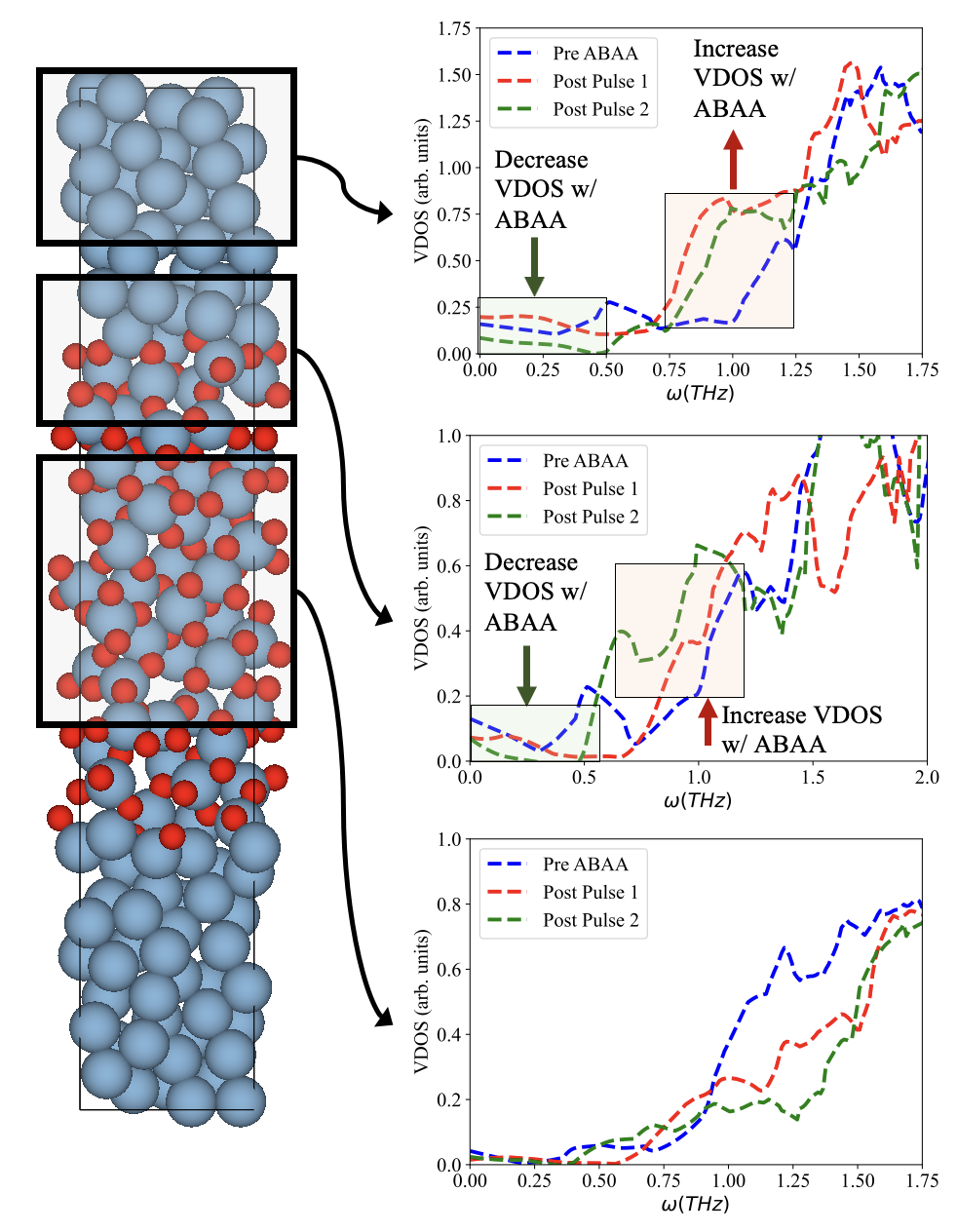}
    \caption{Spatially resolved vibrational density of states in the pure aluminum, aluminum-aluminum oxide interface, and oxide regions of the tunneling barrier before the ABAA protocol (blue lines), as well as after the aging and relaxation process following a single pulse (green lines) or two pulses (red lines). In each region the VDOS below $0.5 $THz is reduced following two pulses. In the aluminum and interface we find this is compensated by an increased VDOS in the vicinity of 1THz. The oxide shows a reduced VDOS across the full examined frequency range.}
    \label{fig:VDOS}
\end{figure}
\section{Tracking the vibrational density of states with machine learned potentials}
A potential source of TLSs in amorphous oxides are defects which give rise to soft-modes localized in real-space, also referred to as vibrons\cite{tyner2025identification}. An example of this is illustrated in Fig.~\eqref{fig:Fig2}. It is therefore important to examine the real-space distribution of the vibrational density of states before and after the ABAA protocol. To make this goal computationally tractable we compute the phonon density of states via Phonopy software package\cite{togo2023first} using the finite difference approach and considering displacements of $0.001 \AA$ on the atoms. To compute forces we utilize a foundational machine-learned potential, namely MACE-MP-0b3, which is optimally suited for this task given the large number of oxides in the training dataset\cite{batatia2022mace,batatia2025foundation}. The computational efficiency of machine-learned interatomic potentials is paramount in opening the possibility to track the vibrational density of states. Computation of the phonon density of states for amorphous oxide tunneling barriers containing hundreds of atoms via traditional density functional perturbation theory would be computationally intractable. This computational expense would be exacerbated by the need to average over multiple molecular-dynamics time steps. By contrast, the MACE machine-learned interatomic potential offers a route to efficiently compute the vibrational density of states over multiple molecular-dynamics frames while maintaining accuracy at the DFT level.
%To verify this, in the supplementary material we present a comparison of the DFT and machine learned phonon density of states for select examples of Al$_{2}$O$_{3}$ and thin film Al. 
\par 
We limit our analysis of the vibrational density of states (VDOS) to the relaxed tunneling barrier (I) before the ABAA protocol, (II) after aging and relaxation following a single pulse, and (III) after aging and relaxation following two alternating pulses. In order to understand the impact of ABAA on the distribution of low-frequency soft-modes which are potential TLS candidates, we spatially resolve the vibrational density of states, considering the oxide, interface, and pure aluminum layers in Fig. \eqref{fig:VDOS}. 
\par 
Examining the data shown in Fig. \eqref{fig:VDOS} it is interesting to note that in each case the minimal VDOS is below $\sim 0.5THz$ is found following application of two alternating pulses. This indicates a reduction in soft-modes which can serve as sources of TLSs. This is also in alignment with the prior finding that total energy is at a minimum following application of two alternating pulses which would suggest enhance stability and a reduction in soft modes. It is further important to note that in the interface and pure aluminum layers the decrease in VDOS below $\sim 0.5 THZ$ is accompanied by an increase in the VDOS near $\sim 1THz$. This suggests that the ABAA protocol may not serve to remove TLSs altogether but rather cause a sift in TLS frequency above the operating range of the qubit. 
\par 
In contrast to the interface and pure aluminum layers, in the bulk oxide we do not observe a clear increase in the VDOS near $\sim 1Thz$ following the ABAA protocol, rather we find an overall decrease in the VDOS in the analyzed range $0-1.75Hz$ following each pulse. This suggests that the ABAA protocol does serve to move the oxide out of a local minima in the potential energy and closer to a global potential energy minimum.

\section{Summary}
The ABAA protocol introduced by Ref. \cite{pappas2024alternating}, has shown efficacy in reduction of TLSs within the oxide tunneling barrier, and more precisely within a frequency range capable of coupling to the qubit. Our work demonstrates that ABAA serves as an effective secondary annealing cycle, allowing the oxide tunneling barrier to escape a local potential minima in the energy landscape and move closer to the global potential energy minima. By traversing the energetic landscape in this manner the density of soft modes, particularly localized vibrational modes below $0.5 THz$ is found to be reduced. This is in agreement with past work proposing TLSs as localized soft modes. This work provides further evidence that such TLSs are not removed entirely, rather the total state count must be preserved, but computations suggest the possibility that the TLS frequency is shifted to higher frequencies outside the range at which a TLS can effectively couple to the qubit. Future work to examine variations in temperature, magnitude of the applied bias and number of applied alternating biases requires an extensive computational effort, but is a promising direction for optimization of the experimental ABAA protocol. 

\par 
\acknowledgments{} 
We are grateful to P. Alpay, J. Cole,  D. Pappas, Y. Rosen, J. Mutus and X. Wang for useful discussions.  AB and AT were supported by Rigetti collaboration USAFOSR Award FA9550-25-1-0103. AT was also supported by European Research Council under the European Union Seventh Framework ERS-2018-SYG 810451 HERO and by Knut and Alice Wallenberg Foundation Grant No. KAW 2019.0068. The computations were enabled by resources provided by the National Academic Infrastructure for Supercomputing in Sweden (NAISS), partially funded by the Swedish Research Council through grant agreement no. 2022-06725. 

\bibliographystyle{apsrev4-1}
\nocite{apsrev41Control}
\bibliography{ref.bib}
\end{document}